\RequirePackage{fix-cm}
\documentclass[smallextended]{svjour3}       
\smartqed  
\usepackage{amsmath}
\usepackage{amssymb,xspace,intmacros,subfigure,amsfonts}
\usepackage{epsfig,epstopdf,graphicx}
\usepackage[title]{appendix}
\usepackage{color}

\newcommand{\xb}{{\textbf{x}}}
\newcommand{\yb}{{\textbf{y}}}

\newcommand{\Cb}{{\textbf{C}}}

\newcommand{\zb}{{\textbf{z}}}

\begin{document}

\title{One Bit Spectrum Sensing in Cognitive Radio Sensor Networks
}


\author{Hadi Zayyani         \and
        Farzan Haddadi \and Mehdi Korki 
}


\institute{Hadi Zayyani \at
              \email{zayyani2009@gmail.com}           
           \and
           Farzan Haddadi \at
              Iran University of Science and Technology
          \and
          Mehdi Korki \at
          Swinburne University of Technology
}

\date{Received: date / Accepted: date}

\maketitle

\begin{abstract}
This paper proposes a spectrum sensing algorithm from one bit measurements in a cognitive radio sensor network. A likelihood
ratio test (LRT) for the one bit spectrum sensing problem is derived. Different from the one bit spectrum sensing research work in the literature, the
signal is assumed to be a discrete random correlated Gaussian process, where the correlation is only available within immediate successive
samples of the received signal. The employed model facilitates the design of a powerful detection criteria with measurable analytical performance.
One bit spectrum sensing criterion is derived for one sensor which is then generalized to multiple sensors. Performance of the detector is analyzed by obtaining closed-form formulas for the probability of false alarm and the probability of detection. Simulation results corroborate the theoretical findings and confirm the efficacy of the
proposed detector in the context of highly correlated signals and large number of sensors.
\end{abstract}

\section{Introduction}
\label{intro}
Cognitive radio (CR) \cite{Mitola99} is an emerging technology to improve the spectrum access 
in wireless sensor networks. It allows unlicensed or secondary users (SUs) to detect and
access any available radio spectrum unused by licensed or primary users (PUs) without
causing harmful interference to PUs. Hence, Spectrum sensing \cite{Ali17x}--\nocite{Quan09}\nocite{Moghimi11}
\nocite{Qing15}\nocite{Khani17}\nocite{Mac17}\nocite{Ahmad17}\nocite{Chen18}\nocite{Wang18}\cite{Sriharipriya18} is a vital component of a CR system to identify the state of the PUs in the network.

Spectrum sensing techniques can be categorized into narrowband spectrum sensing and wideband spectrum sensing techniques \cite{Ali17}.
Narrowband spectrum sensing \cite{Larsson08} investigates the problem of identifying whether a particular slice of the spectrum is idle or not. In contrast, wideband 
spectrum sensing \cite{Sun13} aims to classify individual slices of a wide frequency range, i.e., megaherts (MHz) to gigahertz (GHz) range, to be either vacant or occupied. Therefore, in majority of 
existing wideband spectrum sensing techniques, a simple approach is to acquire the wideband signal samples using a standard Analog to Digital converter (ADC) and then utilize 
appropriate signal processing techniques to identify spectral opportunities. In these techniques, however, the samples of the signal should follow Shannon's theorem: the sampling rate must be at 
least twice the maximum available frequency in the signal, i.e., Nyquist rate, to avoid spectral aliasing. Hence, employing these wideband spectrum sensing techniques results in long sensing delays
or leads to higher computational complexities and hardware costs. As a result, these techniques are inappropriate for a cognitive wireless sensor network (CWSN) \cite{Zhang18}, \cite{Valehi17} 
with simple and affordable sensors. 

A number of techniques have been proposed in the literature to address the challenges, including multiband sensing (FFT-based sensing), wavelet-based sensing, and 
filter-bank sensing \cite{Ali17}. However, these approaches still suffer from the practical issues such as power consumption, feasibility of ultra high sampling ADCs, sensing time and complexity. To 
avoid the high sampling rate or high implementation complexity in Nyquist systems, sub-Nyquist approaches have gained more attention, in which the sampling rates lower than Nyquist rate is 
employed to detect spectral opportunities. One of these sub-Nyquist approaches is compressive sensing (CS) \cite{Dono06}, \cite{CandT06} for detection of sparse signals \cite{ZayyH16} or 
spectrum sensing in cognitive radio framework \cite{Salah18}. However, there are some limitations on CS techniques. For instance, the sensing matrix should be properly selected to satisfy 
some constraints (e.g., nearly orthonormal matrices). Further, the spectrum reconstruction part of CS approach is challenging \cite{Ali17}. 

To simplify the implementation of high sampling ADCs, it is preferred to use low precision ADCs. The extreme case is to use one bit ADCs utilizing sign measurement by a
simple comparator \cite{Ali16}, \cite{Ali18}, \cite{Stein18}. In \cite{Ali16}, an ultra low power wideband spectrum sensing architecture is suggested by utilizing a one bit quantization at
the cognitive radio receiver. In \cite{Ali18}, the same authors used a window-based autocorrelation to provide the power spectral density of the quantized signal. Recently, the authors in
\cite{Stein18} considered the problem of detecting the presence or absence of a random wireless source with minimum latency utilizing an array of radio one bit sensors. 

\subsection{Contribution}
In this paper, a likelihood ratio test (LRT) detector is derived for detection of a random source with one bit measurements. Unlike the above-mentioned research work on one bit spectrum sensing, to reduce the complexity of the one-bit model likelihood, we employ a correlated Gaussian random process for the received signal model, where the correlation is only available within the immediate successive samples of the received signal. The employed model enables use to design a powerful detection criteria with measurable analytical performance.

The detector performance is investigated in single sensor and multiple sensors scenarios. Then, theoretical analysis
of the detector is performed by calculating closed-form formulas for the probability of detection and probability of false alarm. Simulation results show
efficacy of the LRT detector and agreement between experimental and theoretical results. The proposed one bit spectrum sensing detector provides competitive capabilities to save the hardware, power and computing resources by minimizing the ADC output resolution for a large number of sensors in multiple sensor scenarios.

The rest of the paper is organized as follows. Section~\ref{sec:
single} introduces the model, the LRT detector and the theoretical
analysis for the single sensor case. In section~\ref{sec:
multiple}, the same steps are performed in the case of multiple
sensors. Simulation results are presented in Section~\ref{sec:
Sim}. Finally, conclusions are drawn in
Section~\ref{sec: con}.


\section{One bit spectrum sensing: single sensor case}
\label{sec: single}

Consider a random signal $s_i$ for $1\leqslant i \leqslant n$ in
which $n$ is the total number of samples. One bit measurements of
the single sensor is modeled as
\begin{align*}
\mathrm{H_0}:\quad y_i &= \mathrm{sgn}(w_i),\\
\mathrm{H_1}:\quad y_i &= \mathrm{sgn}(s_i+w_i),\quad i=1,2,...,n
\end{align*}
where $w_i$ is Gaussian noise with zero mean and variance
$\sigma^2$, $H_0$ and $H_1$ are hypotheses of absence and presence
of the signal, respectively, and $\mathrm{sgn}(x)$ is the
indicator function ($\mathrm{sgn}(x)=1$ for $x \geqslant 0$ and
$\mathrm{sgn}(x)=0$ for $x<0$). Signal is assumed to be a
correlated Gaussian random process with a covariance matrix which
is toeplitz and banded with bandwidth 3. This means that
correlation is present only between immediate successive samples.
This is the case when sampling rate is less than or equal to twice
the symbol rate of a digitally modulated signal 
Hence, we have $\mathbb E (s^2_i) = \sigma^2_s$
and $\mathbb E (s_i s_{i+1})=r$ while $\mathbb E (s_i s_{i+k} ) =
0$ for $|k|>1$. The problem is to decide the true hypothesis
(absence or presence of the signal) from one bit measurements
$\yb=[\, y_1, y_2, ..., y_n]^\textrm T$.

The Neyman-Pearson LRT detector is defined as \cite{Kaydet}:
\begin{equation}
\label{eq: det} \Lambda_{\mathrm{LR}}=\frac{\mathbb
P(\yb|H_0)}{\mathbb P(\yb|H_1)} \gtrless \! \! \! \! \!
\mathrel{\ooalign{\raisebox{2.0ex}{$\scriptstyle
H_0$}\cr\raisebox{-2.0ex}{$\scriptstyle H_1$}}} \lambda
\end{equation}
where $\mathbb P(\cdot)$ is the probability mass function or
probability depending on the context and $\lambda$ is the
detector's threshold. The likelihood under hypothesis $H_0$ is
equal to $\mathbb P(\yb|H_0)=(\frac{1}{2})^n$. The likelihood
$\mathbb P(\yb|H_1)$ is equal to
\begin{equation*}
\mathbb P(y_1|H_1) \mathbb P(y_2|y_1,H_1) \mathbb
P(y_3|y_2,H_1)... \mathbb P(y_n|y_{n-1},H_1)
\end{equation*}
since we have only one sample dependence between measurements.
Also, we have $ \mathbb P(y_1|H_1)=\frac{1}{2}$. In
Appendix~\ref{app1}, the probability $\mathbb P(y_2|y_1,H_1)$ is
calculated to be
\begin{equation}
\label{eq: fy} \mathbb P(y_2|y_1,H_1)=p^{\, \mathbb I(y_1=y_2)}
(1-p)^ { \mathbb I( y_1 \neq y_2)}
\end{equation}
where
\begin{equation}
\mathbb I(y_1=y_2)=\Big\{\begin{array}{cc}
                    1 & y_1=y_2, \\
                    0 & \text{else}
                  \end{array}
\end{equation}
and
\begin{equation}
p \mathrel{ \mathop :} = \mathbb P(y_2=1|y_1=1,H_1)
\end{equation}

A similar approach shows that $\mathbb P(y_{k+1}|y_{k},H_1) =
p^{\, \mathbb I(y_{k} = y_{k+1})} (1-p)^ {\mathbb I(y_k \neq
y_{k+1})}$ for $2 \leqslant k \leqslant n-1$. Replacing these
conditional probabilities in logarithm of (\ref{eq: det}) followed
by straightforward calculations lead to the final detection
criterion:

\begin{equation}
\label{eq: finaldet1} \sum_{i=1}^{n-1} \mathbb I(\, y_i = y_{i+1})
\gtrless \! \! \! \! \!
\mathrel{\ooalign{\raisebox{2.0ex}{$\scriptstyle
H_1$}\cr\raisebox{-2.0ex}{$\scriptstyle H_0$}}} \eta
\end{equation}
In the derivation, it is assumed that $\ln \frac{p}{1-p}>0$ which
is equivalent to $p>\frac{1}{2}$ or $r>0$. For the case of
$p<\frac{1}{2}$ or equivalently $r<0$, the detection criterion has
the reverse direction. For the case of $r=0$ in which the source
samples like the noise samples are uncorrelated Gaussian random
variables, the energy detector is the sole choice \cite{Kaydet},
\cite{Poor}. Although, one bit measurements have no amplitude
information, so there is no information to decide between the
presence and absence of the signal.

In the following, detection probability and false alarm
probability are calculated for the LRT detector of (\ref{eq:
finaldet1}) in the case of $p>\frac{1}{2}$. Detection statistic is
defined as $Y=\sum_{i=1}^{n-1} \mathbb I( \, y_i = y_{i+1})$, The
decision is
\begin{equation}
\hat{d}=\Big\{\begin{array}{cc}
                    1 & Y \geqslant \eta \\
                    0 & Y< \eta.
                  \end{array}
\end{equation}
Hence, the false alarm probability $P_{\mathrm{fa}} = \mathbb
P(\hat{d}=1|H_0) = \mathbb P( \, Y > \eta |H_0)$ is equal to
\begin{equation}
P_{\mathrm{fa}} = \mathbb P \Bigg( \sum_{i=1}^{n-1} \mathbb I( y_i
= y_{i+1}) \geqslant \eta \, | \, H_0 \Bigg) \approx Q \Big(\frac{\eta
-\mu_0} {\sigma_0} \Big)
\end{equation}
where $Q(\cdot)$ is the Q-function, $\mu_0 = E(Y|H_0)$,
$\sigma^2_0$ is the variance of $Y$ subject to hypothesis $H_0$
and it is assumed that the detection statistic $Y=\sum_ {i=1}^
{n-1} \mathbb I( y_i = y_{i+1})$ is Gaussian due to the Central
Limit Theorem (CLT). In Appendix~\ref{app2}, $\mu_0$ and
$\sigma^2_0$ are calculated to be:
\begin{equation*}
\mu_0=\frac{1}{2}(n-1)
\end{equation*}
\begin{equation}
\label{eq: sigmaH01} \sigma^2_0= \frac{1}{4}(n-1)
\end{equation}

The detection probability $P_\textrm d = \mathbb P(\hat{d}=1|H_1)
= \mathbb P(Y \geqslant \eta \, | \, H_1)$, we have:
\begin{equation}
P_{\mathrm{d}} = \mathbb P \Bigg( \sum_{i=1}^{n-1} \mathbb I( y_i
= y_{i+1}) \geqslant \eta \, | \, H_1 \Bigg) \approx Q \Big(\frac{\eta
-\mu_1} {\sigma_1} \Big)
\end{equation}
where $\mu_1 = \mathbb E(Y|H_1)$, $\sigma^2_1$ is the variance of
$Y$ subject to hypothesis $H_1$. In Appendix~\ref{app3}, $\mu_1$
and $\sigma^2_1$ are calculated as:
\begin{eqnarray}
\mu_1 &=& 2p(n-1) \\
 \label{eq: sigmaH11} \sigma^2_1 &=& 2p(1-2p)(n-1).
\end{eqnarray}

\section{One bit spectrum sensing: sensor network case}
\label{sec: multiple} Consider a sensor network with $N$ nodes.
Each sensor performs a one bit measurement as
\begin{align*}
\mathrm{H_0}:\quad y_{ki} &= \mathrm{sgn}(w_{ki}),\\
\mathrm{H_1}:\quad y_{ki} &= \mathrm{sgn}(s_i + w_{ki})
\end{align*}
where $1\leqslant i\leqslant n$ is the time index, $1\leqslant
k\leqslant N$ is the sensor index, $N$ is the total number of
sensors, $w_{ki}$ is the Gaussian noise of $k$'th sensor, and
$s_i$ is the signal sample with the same model as assumed in
section~\ref{sec: single}.

The LRT detector will be \cite{Kaydet}:
\begin{equation}
\label{eq: det1} \Lambda_{\mathrm{LR}}=\frac{ \mathbb
P(\xb_1,\xb_2,...,\xb_n|H_0)} { \mathbb P(\xb_1, \xb_2,...,
\xb_n|H_1)} \gtrless \! \! \! \! \!
\mathrel{\ooalign{\raisebox{2.0ex}{$\scriptstyle
H_0$}\cr\raisebox{-2.0ex}{$\scriptstyle H_1$}}} \lambda
\end{equation}
where $\xb_i=[y_{1i} \quad y_{2i} \quad ... \quad y_{Ni}]^ \textrm
T$ is the measurements of all sensors at $i$'th time instant. We
will have $ \mathbb P( \xb_1, \xb_2, ..., \xb_n | H_0) = (
\frac{1}{2})^{nN}$. Also, $\mathbb P(\xb_1, \xb_2,..., \xb_n|H_1)$
is equal to:
\begin{equation}
\label{eq: fx} \mathbb P(\xb_1|H_1) \mathbb P(\xb_2|\xb_1,H_1)
\mathbb P(\xb_3|\xb_2,H_1)... \mathbb P(\xb_n|\xb_{n-1},H_1)
\end{equation}
where $\mathbb P(\xb_1|H_1)=(\frac{1}{2})^{N}$. Also $\mathbb
P(\xb_2|\xb_1,H_1)$ is equal to $ \prod_{k=1}^N \mathbb
P(y_{k2}|y_{k1},H_1)$ where $\mathbb P(y_{k2}|y_{k1},H_1) = p^{\,
\mathbb I(y_{k2} = y_{k1})} (1-p)^{ \mathbb I(y_{k2}
 \neq y_{k1})}$. We will have

\begin{equation}
\mathbb P(\xb_2|\xb_1,H_1) = (1-p)^N \Big( \frac{p}{1-p} \Big)
^{\sum_k \mathbb I(y_{k2} = y_{k1})}
\end{equation}

Similar calculations lead to $\mathbb P(\xb_{i+1}|\xb_{i},H_1) =
(1-p)^N (\frac{p}{1-p})^ {\sum_k \mathbb I(y_{k,i+1} = y_{ki})}$.
Replacing these conditional probabilities into (\ref{eq: fx}) and
(\ref{eq: det1}), leads to the following criterion for
$p>\frac{1}{2}$:
\begin{equation}
\label{eq: finaldet} Y \mathrel{\mathop :} =
\sum_{i=1}^{n-1}\sum_{k=1}^{N} \mathbb I(y_{k,i+1} = y_{ki})
\gtrless \! \! \! \! \!
\mathrel{\ooalign{\raisebox{2.0ex}{$\scriptstyle
H_1$}\cr\raisebox{-2.0ex}{$\scriptstyle H_0$}}} \eta
\end{equation}
which is a direct generalization of detection criterion in single
sensor case in (\ref{eq: finaldet1}). For the case of
$p<\frac{1}{2}$, the direction of the detection criterion in
(\ref{eq: finaldet}) is reversed. The case of $p=\frac{1}{2}$ is
the same as that in the single sensor case where there is no
information to detect the presence of the signal.

In the next step,  detection probability and  false alarm
probability are calculated for the LRT detector of (\ref{eq:
finaldet}) when $p>\frac{1}{2}$. False alarm probability
$P_{\mathrm{fa}}= \mathbb P(\hat{d}=1|H_0)= \mathbb P(Y> \eta
|H_0)$ is equal to
\begin{equation}
P_{\mathrm{fa}} = Q \Big(\frac{ \eta - m_0} {s_0} \Big)
\end{equation}
where $m_0 = \mathbb E(Y|H_0)$ and $s^2_0$ is the variance of $Y$
subject to hypothesis $H_0$ and it is assumed that the detection
statistic $Y$ is Gaussian due to the CLT. In Appendix~\ref{app4},
$m_0$ and $s^2_0$ are calculated to be:
\begin{equation*}
m_0 = \frac{1}{2}(n-1)N
\end{equation*}
\begin{equation}
\label{eq: sigmaH02} s^2_0 = \frac{1}{4}(n-1)N
\end{equation}

The detection probability $P_\textrm d = \mathbb P(\hat{d}=1|H_1
)= \mathbb P(Y\geqslant \eta|H_1)$. we have:
\begin{equation}
P_\textrm d = Q \Big(\frac{\eta-m_1}{s_1} \Big)
\end{equation}
where $m_1 = \mathbb E(Y|H_1)$ and $s^2_1$ is the variance of $Y$
subject to hypothesis $H_1$. In Appendix~\ref{app5}, $m_1$ and
$s^2_1$ are calculated as
\begin{eqnarray}
m_1 &=& 2p(n-1)N \\
\label{eq: sigmaH12} s^2_1 &=& 2p(1-2p)(n-1)N
\end{eqnarray}

\section{Simulation Results}
\label{sec: Sim} This section presents the simulation results.
Correlated random signal is generated as described in
section~\ref{sec: single}, with parameters $r$ and $\sigma_s=1$.
Noise is generated as a Gaussian uncorrelated random process with
zero mean and variance $\sigma=10^{-2}$. In all simulations,
number of time samples are assumed to be $n=20$. For comparison of
the detectors, the detection probability versus false alarm
probability known as Receiver Operating Characteristic (ROC) is
depicted. For the monte carlo simulation, the experiments are
repeated  20000 times and  detection probability and  false alarm
probability are averaged over all the trials. Moreover, to verify
the theoretical analysis, we compared the experimental results
with the theoretical results. Two experiments are performed for
single sensor and multiple sensor cases.

\begin{figure}[tb]
\begin{center}
\includegraphics[width=9cm]{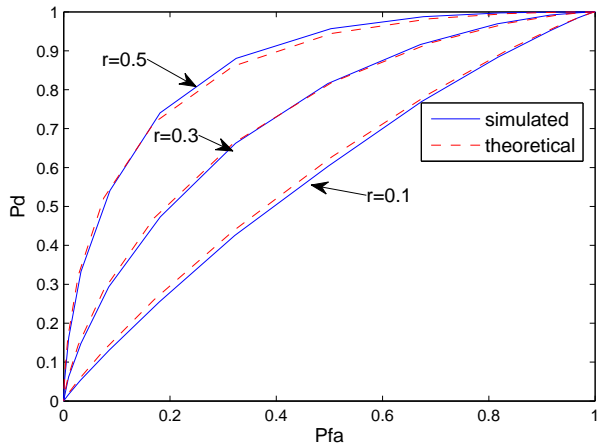}
\end{center}
\caption{ROC curve of the LRT detector for single sensor case.}
\label{fig2}
\end{figure}

In the first experiment, a single sensor is used for spectrum
sensing. Four signals are examined with correlation coefficients
$r=0.1$, $0.3$, and $0.5$. A good agreement between experimental and theoretical ROC
curves are shown in Fig~\ref{fig2}. Also,
it shows that by increasing the correlation coefficient, the
performance of the detector improves.

\begin{figure}[tb]
\begin{center}
\includegraphics[width=9cm]{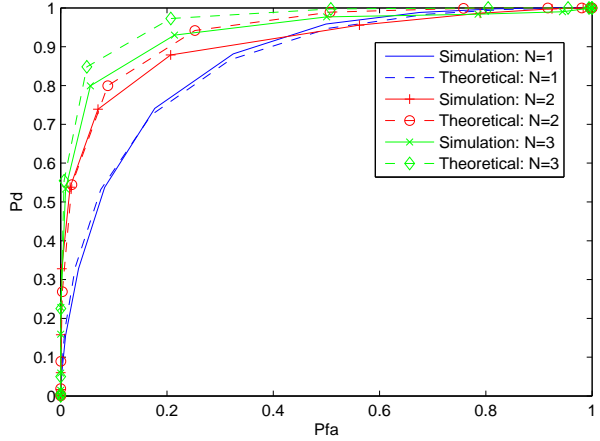}
\end{center}
\caption{ROC curve of the LRT detector for multiple sensor case.}
\label{fig3}
\end{figure}

In the second experiment, we utilize a cognitive sensor network
with $1$, $2$, and $3$ sensors. The correlation coefficient of
signal is $r=0.5$. ROC curves are sketched in Fig.~\ref{fig3}. It
shows that by increasing the number of sensors, the detector
performance improves. It also demonstrates a good agreement
between theoretical and experimental ROC curves.

\section{Conclusion}
\label{sec: con} In this paper, we have derived the LRT detector
for one bit spectrum sensing problem in single sensor and multiple
sensor cases for a correlated Gaussian signal model. The detectors utilize
correlation available within successive samples of the received signal to obtain the detection criteria. Closed-form detection and false alarm
probabilities are derived in single and multiple sensor scenarios.
Simulation results show the efficacy of the detector specially
when the correlation coefficient is large or the number of sensors
increases. Moreover, the simulations results corroborate the theoretical analysis. The proposed one bit spectrum sensing detector provides competitive capabilities to save the hardware, power and computing resources by minimizing the ADC output resolution for a large number of sensors in multiple sensor scenarios.

\begin{appendices}
\section{Calculating $\mathbb P(y_2|y_1,H_1)$}
\label{app1} We first calculate the four probabilities $\mathbb
P(y_2=1|y_1=1,H_1)= \mathrel{\mathop :} p$ , $ \mathbb
P(y_2=0|y_1=1,H_1)=1-p$ , $\mathbb P(y_2=1|y_1=0,H_1)=p\,'$ , and
$ \mathbb P(y_2=0|y_1=0,H_1)=1-p\,'$. The probability $p= \mathbb
P (y_2=1|y_1=1,H_1)$ is equal to

\begin{equation*}
p = \mathbb P(w_2+s_2\geqslant 0|w_1+s_1\geqslant 0)=
\end{equation*}
\begin{equation}
\frac{\mathbb P(z_1 \geqslant 0, z_2 \geqslant 0)}{\mathbb P(z_1
\geqslant 0)} = 2 \, \mathbb P(z_1 \geqslant 0,z_2 \geqslant 0)
\end{equation}
where $z_1=s_1+w_1$, $z_2=s_2+w_2$ and $p(z_1 \geqslant
0)=\frac{1}{2}$. To calculate $p(z_1 \geqslant 0,z_2 \geqslant
0)$, note that $z_1$ and $z_2$ are correlated Gaussian random
variables  with covariance matrix $\Cb$ with elements
$C_{11}=E(z^2_1)=\sigma^2_s+\sigma^2$, $C_{12}=C_{21}=E(z_1z_2)=r$
and $C_{22}=E(z^2_2)=\sigma^2_s+\sigma^2$. Therefore, joint
probability density function (pdf) is $f(z_1,z_2) = \frac{1}{2\pi
\sqrt{\mathrm{det} (\Cb)}} \exp{(-\frac{1}{2} \zb \Cb^{-1} \zb^
\text T)}$ where $\zb=[z_1 \; z_2]$. Hence, we have
$p=2\int_{0}^{+\infty}\int_{0}^{+\infty}f(z_1,z_2)dz_1dz_2$. To
calculate the other probability $p\,'$, we consider that $p\,'=
\mathbb P(y_2=1 | y_1=0 , H_1) = \frac{\mathbb P(y_2=1 , y_1=0 |
H_1)}{ \mathbb P(y_1=0 | H_1)} = 2 \, \mathbb P(y_2=1 , y_1=0 |
H_1) = 2(\frac{1}{2}- \frac{p}{2})=1-p$ which leads to (\ref{eq:
fy}).

\section{Calculating $\mu_0$ and $\sigma^2_0$}
\label{app2} We have $\mu_0 = \sum_{i=1}^{n-1} \mathbb E \,
\mathbb I(y_i = y_{i+1}|H_0)=\frac{1}{2}(n-1)$. Also, we have
$\sigma^2_0 = \mathbb E(Y^2|H_0) - \mathbb E^2(Y|H_0)$ in which $
\mathbb E(Y|H_0) = \frac{1}{2}(n-1)$ and
\begin{equation}
\label{eq: fourthsigma1} \mathbb E(Y^2|H_0) = \sum_{i,i'} \mathbb
E(\, \mathbb I(y_i = y_{i+1}) \, \mathbb I(y_{i'} = y_{i'+1}) |
H_0)
\end{equation}
where the expectation is equal to
\begin{equation*}
\label{eq: pdel1} \mathbb P( \, (y_i = y_{i+1}) \land (y_{i'} =
y_{i'+1})  | H_0) = \begin{cases}
                    \frac{1}{2} & : \quad i=i' \\
                    \frac{1}{4} & : \quad i \neq i'
                  \end{cases}
\end{equation*}
Replacing (\ref{eq: pdel1}) into (\ref{eq: fourthsigma1}) results
in (\ref{eq: sigmaH01}).

\section{Calculating $\mu_1$ and $\sigma^2_1$}
\label{app3} We have $\mu_1 = \sum_{i=1}^ {n-1} \mathbb E \,
\mathbb I( \, y_i = y_{i+1}|H_1) = \sum_{i=1}^ {n-1} \mathbb P( \,
y_i = y_{i+1} |H_1)=2p(n-1)$. Also, we have $\sigma^2_1 = \mathbb
E(Y^2|H_1) - \mathbb E^2(Y|H_1)$ in which $ \mathbb E(Y|H_1) =
2p(n-1)$ and
\begin{equation}
\label{eq: fourthsigmaa1} E(Y^2|H_1)=\sum_{i,i'} \mathbb E(\,
\mathbb I(y_i = y_{i+1}) \, \mathbb I(y_{i'} = y_{i'+1}) | H_1)
\end{equation}
where the expectation is equal to
\begin{equation}
\label{eq: pdell} \mathbb P( \, (y_i = y_{i+1}) \land (y_{i'} =
y_{i'+1})  | H_1) = \quad \quad \quad \quad \quad \quad \quad
\end{equation}
\begin{equation*}
\begin{cases}
\mathbb P( \, y_i = y_{i+1}  | H_1) = 2p &  : \quad i=i' \\
\mathbb P( \, y_i = y_{i+1}  | H_1)\mathbb P( \, y_{i'} = y_{i'+1}
| H_1) = 4p^2 & : \quad i \neq i'
\end{cases}
\end{equation*}
Replacing (\ref{eq: pdell}) into (\ref{eq: fourthsigmaa1}) results
in (\ref{eq: sigmaH11}).

\section{Calculating $m_0$ and $s^2_0$}
\label{app4} We have $m_0= \sum_{i=1}^ {n-1} \sum_{k=1}^{N}
\mathbb E \, \mathbb I(y_{ki} = y_{k,i+1} | H_0)  =
\frac{1}{2}(n-1)N$. Also, we have $s^2_0 = \mathbb E (Y^2| H_0) -
\mathbb E^2(Y| H_0)$ in which $ \mathbb E(Y|H_0) = \frac{1}{2}
(n-1)N$ and
\begin{equation}
\label{eq: fourthsigma2} \mathbb E(Y^2|H_0) = \!\!\!\!
\sum_{i,k,i'\! ,k'} \!\!\! \mathbb E(\, \mathbb I(y_{ki} =
y_{k,i+1}) \, \mathbb I(y_{k'i'} = y_{k',i'+1}) | H_0)
\end{equation}
where the expectation is equal to
\begin{equation}
\label{eq: pdel2} \mathbb P( \, (y_{ki} = y_{k,i+1}) \land
(y_{k'i'} = y_{k',i'+1})  | H_0) =
\end{equation}
\begin{equation*}
\begin{cases}
                    \frac{1}{2} & : \quad i=i' \land k = k' \\
                    \frac{1}{4} & : \quad i \neq i' \lor k \neq k'
\end{cases}
\end{equation*}
Replacing (\ref{eq: pdel2}) into (\ref{eq: fourthsigma2}) results
in (\ref{eq: sigmaH02}).

\section{Calculating $m_1$ and $s^2_1$}
\label{app5} We have $m_1 = \sum_{i=1}^{n-1} \sum_{k=1}^{N}
\mathbb E \, \mathbb I(y_{ki} = y_{k,i+1} | H_1) = \sum_{i,k}
\mathbb P (y_{ki} = y_{k,i+1} | H_1) = 2p(n-1)N$. Also, we have
$s^2_1 = \mathbb E(Y^2|H_1) - \mathbb E^2(Y|H_1)$ in which $
\mathbb E(Y|H_1) = 2p(n-1)N$ and
\begin{equation}
\label{eq: fourthsigmaa2} \mathbb E(Y^2|H_1) = \!\!\!\!
\sum_{i,k,i',k'} \!\!\! \mathbb E(\, \mathbb I(y_{ki} = y_{k,i+1})
\, \mathbb I(y_{k'i'} = y_{k',i'+1}) | H_1)
\end{equation}
where the expectation is equal to
\begin{equation}
\label{eq: pdell2} \begin{array}{l} \mathbb P( \, (y_{ki} =
y_{k,i+1}) \land
(y_{k'i'} = y_{k',i'+1})  | H_1) = \nonumber  \\
\mathbb I (i=i' \land k = k') \mathbb P( \, y_{ki} = y_{k,i+1}  | H_1) \; +  \\
\mathbb I (i \neq i'  \lor k \neq k') \mathbb P( \, y_{ki} =
y_{k,i+1}  | H_1) \mathbb P( \, y_{k'i'} = y_{k',i'+1} | H_1)
\end{array}
\end{equation}
which is
\begin{equation}
\label{eq: pdell2} \mathbb P( \, (y_{ki} = y_{k,i+1}) \land
(y_{k'i'} = y_{k',i'+1})  | H_1)
\end{equation}
\begin{equation*}
 = \begin{cases}
                    2p & : \quad i=i' \land k=k' \\
                    4p^2 & : \quad i \neq i' \lor k \neq k'
    \end{cases}
\end{equation*}

Replacing (\ref{eq: pdell2}) into (\ref{eq: fourthsigmaa2})
results in (\ref{eq: sigmaH12}).

\end{appendices}

\end{document}